\documentstyle[a4,12pt,epsf,axodraw]{article}  

\begin{document} 
 
\title{{\bf Inclusive meson production in peripheral  
collisions of ultrarelativistic heavy ions }} 
 
\author{
K.A.Chikin\footnote{E-mail: const\symbol{64}lav1.npi.msu.su},  
V.L.Korotkih\footnote{E-mail: vlk\symbol{64}lav1.npi.msu.su}, 
A.P.Kryukov\footnote{E-mail: kryukov\symbol{64}theory.npi.msu.su},  
L.I.Sarycheva\footnote{E-mail: lis\symbol{64}alex.npi.msu.su} \\ 
{\em Institute of Nuclear Physics, Moscow State University}\\ \bigskip 
{\em 119899 Moscow, Russia} \\
I.A.Pshenichnov\footnote{E-mail: pshenichnov\symbol{64}nbi.dk}\\
{\em Institute for Nuclear Research, Russian Academy of Science}\\ \bigskip 
{\em 117312 Moscow, Russia }\\
J.P.Bondorf\footnote{E-mail: bondorf\symbol{64}nbi.dk},
I.N.Mishustin\footnote{E-mail: mishustin\symbol{64}nbi.dk} \\ 
{\em Niels Bohr Institute, DK-2100 Copenhagen, Denmark }\\ 
}\date{} 
\maketitle
\begin{abstract}
There exist several proposals to use Weizs\"{a}cker-Williams photons generated
by ultrarelativistic heavy ions to produce exotic particles in 
$\gamma\gamma$ fusion reactions. To estimate the background 
conditions for such reactions
we analyze various mechanisms of meson production in very peripheral
collisions of ultrarelativistic heavy ions at RHIC and LHC energies.
Besides $\gamma\gamma$ fusion they include also electromagnetic $\gamma A$
interactions and strong nucleon-nucleon interactions in grazing $AA$
collisions.  All these processes are characterised by low multiplicities
of produced particles.  $\gamma A$ and $AA$ events are simulated by
corresponding Monte Carlo codes, RELDIS and FRITIOF.  In each of these
processes a certain fraction of pions is produced close to the mid-rapidity
region that gives a background for the $\gamma\gamma$ events.  The
possibility of selecting mesons produced in $\gamma\gamma$ fusion events
via different $p_t$ cut procedures is demonstrated.
\end{abstract}

\section{Introduction} 
 
According to the impact parameter, $b$, different phenomena 
take place in collisions of ultrarelativistic heavy ions.  
They can be divided into the following three categories. 
 
Central collisions $(b\approx 0)$, i.e. the collisions with nearly full 
nuclear overlap  fall into the first category. Such collisions provide  
conditions for the creation of very hot and dense nuclear matter. 
It is the aim of future experimental programs at  
Large Hadron Collider (LHC) at CERN~\cite{LHC} and   
Relativistic Heavy Ion Collider (RHIC) at 
Brookhaven National Laboratory~\cite{RHIC} to study a possible phase  
transition of nuclear and hadronic  
matter into the so-called quark-gluon plasma at high energy  densities.  
As it is expected, the quark-gluon plasma can be created under such extreme 
conditions which are similar to those existed in the Early Universe 
soon after the Big Bang. 
 
The second category contains collisions with partial overlap of 
nuclei ($b<R_1+R_2$, $R_1$ and $R_2$ are the nuclear radii). 
In such collisions the residual spectators  
remain relatively cold.  
Short-range interaction via strong nuclear forces is restricted mainly to  
the participant zone. In the whole  
set of minimum-bias events the number of peripheral nuclear collisions is 
significant due to the geometrical factor $2\pi b$.  
The general picture of ultrarelativistic heavy-ion collisions at LHC and 
RHIC will be incomplete without considering such peripheral 
collisions.  
Non-central heavy-ion collisions are considered as a place to look for 
the disoriented chiral condensates~\cite{Asakawa} and elliptic 
flow~\cite{Kolb}.  
Combining Hanbury-Brown-Twiss method with a determination of the reaction  
plane~\cite{Heiselberg} one is able to study the size, deformation and 
opacities of the particle emission sources.    
With the aim to study the  
central collisions, one has to know the exact properties of 
complementary peripheral collisions.  A proper  
rejection of these "background" collision events will be crucial in the 
extraction of events in which the quark-gluon plasma may be created. 
 
In the collisions of the third category the impact parameter exceeds the 
sum of nuclear radii 
($b>R_1+R_2$). Therefore, there is no overlap of nuclear densities, 
but nevertheless, 
one or both nuclei may be disintegrated by the long-range electromagnetic 
forces. This process of Electromagnetic Dissociation (ED) is a well-known  
phenomenon~\cite{BertBaur,Krauss,BaurHen}. The interaction can be treated 
in terms of equivalent photons representing the Lorentz-boosted 
Coulomb field of heavy ions. At ultrarelativistic energies the  
ED cross section exceeds considerably the pure nuclear cross section for  
heavy colliding nuclei. This fact was confirmed recently in  
experiments~\cite{Datz}. The electromagnetic collision events are less  
violent then the collisions with nuclear interactions. Thus the  
average particle multiplicities are essentially  
lower~\cite{Pshenichnov, Pshenichnov2} and 
the main part of nucleons and mesons is produced in the regions of 
projectile and target fragmentation, very far from the mid-rapidity region.   
 
Besides the action of virtual (equivalent) photons on colliding 
nuclei, the photons from two nuclei can collide (fuse) and
produce various secondary particles. Photon-photon 
($\gamma\gamma$) physics may be investigated in such  
collisions~\cite{Krauss,BaurHen,CMS}. The idea to produce exotic  
particles via $\gamma\gamma$ fusion in heavy-ion colliders has  
been put forward more then 10 years ago in Ref.~\cite{Grabiak}. 
Many authors have further developed this idea since 
that time. The production of different particles from $\mu^{\pm}$,  
$\tau^{\pm}$ leptons to Higgs bosons and supersymmetric particles has 
been considered. The full list of references can be found in the 
recent reviews~\cite{Krauss,BaurHen}.    
 
As one can see, there exist several mechanisms of particle production 
in peripheral collisions of ultrarelativistic heavy ions: 
$\gamma\gamma$ fusion, $\gamma A$ interaction and grazing nuclear collisions  
(see Figs.~\ref{NNproc1}--\ref{NNproc3}).  
In the present paper we study the production of 
$\pi^\pm$ and  $\pi^0$  mesons via these mechanisms.   
We discuss the general features of pion production and calculate the 
contributions of different mechanisms. Prior to investigating  
exotic particle  production in $\gamma\gamma$ collisions a certain 
``calibra\-tion'' is necessary for the theoretical methods and  
experimental techniques.  
  
Similar investigations have been performed specially for STAR  
detector at RHIC~\cite{Klein,Nystrand} and corresponding acceptance cuts  
were applied from the beginning. The CMS detector which will be installed 
at LHC~\cite{CMSbook} can also be used for studying the meson production 
in peripheral collisions. Even having in mind a plan to use another  
experimental set-up for such studies one can estimate first   
$\gamma\gamma$ signal to background ratio without any  
acceptance cuts.  Such comparison of different mechanisms provides a 
guide-line which is free of limitations and restrictions of existing 
experimental facilities. This is useful for possible extensions and updates
of existing detectors. 
  
In the present paper we pay the main attention to the inclusive cross 
sections of meson production by different mechanisms.  
This study is complementary to a very recent one~\cite{Klein2},  
which deals with the exclusive meson production channels.  
As we expect, in order to select 
the rare events with the single produced vector meson~\cite{Klein2}, 
one has to reject the background due to meson production  
($\pi^+$, $\pi^-$, $\pi^0$) via  
$\gamma A$ process.  This background may be estimated in the framework of 
the models used in the present paper.  
 
A meson production event may be followed by a partial disintegration of a  
nucleus, i.e. a branchy process may take place. We study such reactions  
in addition to the coherent meson production process considered in  
Ref.~\cite{Klein2} when the colliding nuclei remain intact. Different  
characteristics of pion production by the equivalent photons are calculated  
by the RELDIS code~\cite{Pshenichnov2} based on the extended model of 
photonuclear reactions~\cite{Iljinov}. The FRITIOF Monte Carlo event 
generator, version 7.1~\cite{FRITIOF} is used to study the properties of 
grazing nuclear collisions.  

\newpage
\section{Distant electromagnetic collisions} 
\subsection{Equivalent photon spectra} 
Let us consider the point-like charge $Ze$ moving with  velocity $v$ at  
the impact parameter $b$. In the Weizs\"{a}cker-Williams  
approximation~\cite{BertBaur,Krauss,BaurHen}  
the spectrum of equivalent photons is given by: 
\begin{equation} 
N(\omega,b)=\frac{Z^2 \alpha}{\pi^2 b^2 \beta^2} x^2  
   \left( K_1^2(x)+\frac{1}{\gamma^2}K_0^2(x) \right) 
\label{aspectr} 
\end{equation} 
where $x=\omega b/\gamma v$, $\beta =v/c$ and $\gamma =(1-\beta^2)^{-1/2}$ is  
the Lorentz factor of the moving charge.  
$K_0$ and $K_1$ are the modified Bessel functions of zero and first order. 
 
The total number of photons with energies $\omega_1$ and  
$\omega_2$ colliding at the point $P$ is obtained by the  
integration of the equivalent photon spectra over  
the distances $b_1$ and $b_2$~\cite{BaurFilho} (see Fig.~\ref{ephoton}): 
\begin{equation} 
F(\omega_1,\omega_2)= 
2 \pi \int_{R_1}^{\infty} b_1 db_1  \int_{R_2}^{\infty} b_2 db_2  
\int_0^{2 \pi} d \phi 
N(\omega_1,b_1) N(\omega_2,b_2) \Theta(B^2), 
\label{aaspectr} 
\end{equation} 
where $R_1$ and $R_2$ are the nuclear radii, $\Theta$ is the step function  
and $B^2=b_1^2+b_2^2-2 b_1 b_2 \cos{\phi}-(R_1+R_2)^2$.
 
The $\gamma \gamma$ luminosity calculated in the double equivalent photon 
approximation~\cite{BaurFilho} is: 
$$ 
\frac{d^2L}{dW dy} = 
\frac{2}{W} F(\frac{W}{2} e^{y},\frac{W}{2} e^{-y}), 
$$ 
where the energy squared $W^2=4\omega_1 \omega_2$ and the  
rapidity $y=1/2\,\ln(\omega_1/\omega_2)$  
of the $\gamma \gamma$ system were introduced. 

The transverse momentum of produced single meson turns out to be small 
$p_t\leq 1/R_{1,2}$~\cite{BaurHen,CMS}. Different nuclear charge formfactors 
may be used in calculations~\cite{BaurHen,CMS}, but in any
approximation the $p_t$ values for a produced meson turns out to be less 
then 30 MeV/c in $\rm{PbPb}$ collisions.  
In the following we shell see how this feature may be used to 
disentangle $\gamma\gamma$ events from other processes.

\subsection{Production of a single meson in $\gamma\gamma$ fusion} 
 
The production of a single meson in the $\gamma\gamma$ fusion is the  
simplest process in our consideration. The calculation technique 
within the framework of the Weizs\"{a}cker-Williams formalism  for 
such a process is well-known~\cite{BaurHen,BaurFilho}.   
 
The cross section to produce a meson with mass $M_R$ is given by:  
\begin{equation} 
\sigma=\int {d\omega_1\over \omega_1}\int {d\omega_2\over \omega_2} 
F(\omega_1,\omega_2)\sigma_{\gamma\gamma 
\rightarrow M_R}, 
\end{equation} 
\noindent where the number of colliding photons,  
$F(\omega_1,\omega_2)$, is taken from Eq.~(\ref{aaspectr}). 
 Taking $\Theta=1$ one can simplify
the calculation, since  in this case the integral of 
Eq.~(\ref{aaspectr}) may be
reduced to the product of the Weizs\"{a}cker-Williams spectra,   
$N(\omega)$, integrated 
over the impact parameter~\cite{BertBaur,Krauss,BaurHen}.  
For light mesons, $M_R\ll\gamma/(R_1+R_2)$, the resulting integral is changed 
by several percent only. 
This simplification may be not appropriate for heavy mesons, since they
can only be produced by a pair of photons from the high-energy 
tail of the equivalent photon spectrum. Such mesons are  
produced mainly in close collisions, where $b\approx(R_1+R_2)$
and the region of nuclear overlap cannot be neglected. 

The cross section of elementary process $\gamma\gamma\rightarrow M_R$ may be 
calculated~\cite{Krauss,BaurHen,BaurFilho} as: 
\begin{equation} 
\sigma_{\gamma\gamma \rightarrow M_R}=8\pi^2(2J_R+1)
\Gamma_{M_R\rightarrow \gamma\gamma}\delta(W^2-M_R^2)/M_R, 
\end{equation} 
\noindent  where $J_R$, $M_R$ and $\Gamma_{M_R\rightarrow \gamma\gamma}$ 
are the spin, mass and two-photon decay width of the meson $R$, 
 $W$ is the c.m. energy squared of the colliding photons. 

The corresponding differential cross section for producing a meson with  
mass $M_R$ is given by  
\begin{equation} 
{d\sigma_R \over dy}=8\pi^2(2J_R+1) 
\Gamma_{M_R\rightarrow \gamma\gamma}
F(\frac{M_R}{2}e^y,\frac{M_R}{2}e^{-y})/M_R^3. 
\label{difoneres}
\end{equation}   

The exclusive cross sections for producing a singe meson were calculated 
by many authors. In Ref.~\cite{Krauss} the equivalent photon cross section 
has been derived directly
from the first QED principles. Three types of nuclear formfactors were used in 
the calculations~\cite{Krauss} corresponding to a homogeneously charged 
sphere, a Gaussian-shaped and a point-like charge distributions. 
In this respect the approach of Ref.~\cite{Krauss} is different 
from the more phenomenological method of Ref.~\cite{BaurFilho} 
which we basically follow. Moreover, 
even following the authors of
Ref.~\cite{BaurFilho} one can use different values for the nuclear radii, 
$R_{1,2}$. The real nuclear charge distribution with a diffuse boundary 
should be approximated by a distribution with a sharp boundary that leads 
to some uncertainties. 
A straightforward comparison of the results of different 
authors~\cite{Krauss,BaurHen,BaurFilho}  
seems to be difficult due to different values of $M_R$, $\Gamma_{\gamma\gamma 
\rightarrow M_R}$ and $R_{1,2}$ used in the different papers. 

With the aim to understand the sensitivity of numerical results to the choice
of parameters
we repeated the calculation of the cross sections 
for single meson production in ultrarelativistic heavy-ion collisions, 
see Tabs.~\ref{tab2}--\ref{tab4}. 
We used  $M_R$ and $\Gamma_{M_R\rightarrow \gamma\gamma}$
values according to the corresponding papers~\cite{Krauss,BaurHen,BaurFilho}.
All the cross sections were calculated for the nuclear 
radii $R_{1,2}=1.2A^{1\over3}$.

As one can see from Tabs.~\ref{tab2}--\ref{tab4}, the different approaches 
give quite similar
results for the lightest mesons $\pi^0$, $\eta$. On the contrary, as it was
expected, the difference is noticeable for heavy mesons like 
$\eta_b$, $\eta_c$. 
Nevertheless this situation seems to be acceptable due to the following reasons. 
First, in the present paper we will consider mainly the production of 
the lightest $\pi$ mesons. Second, because of uncertainties in 
Particle Properties Data on $M_R$ and $\Gamma_{M_R\rightarrow\gamma\gamma}$
for heavy mesons~\cite{PartDat}\footnote[1]{This is particularly true 
for scalar mesons $f_0(975)$ and $f_0(1275)$ 
with large decay widths causing a strong overlap of individual 
resonances, see details in Ref.~\protect\cite{PartDat}.} it is difficult 
to predict with confidence the exclusive cross section values for 
such mesons. Therefore, the uncertainties of the method itself
become less important for heavy mesons.

\subsection{Inclusive cross section of $\pi^0$ production in the 
DRP model} 
 
Pions can be produced not only in the direct process 
$\gamma\gamma\rightarrow\pi^0$. 
The $\gamma \gamma $ fusion  produces several unstable heavy mesons 
such as $\eta ,f_0$,... and decay products of these mesons may
contain $\pi^0$'s. As one can see in the following, such two-step processes,
$\gamma\gamma\rightarrow R\rightarrow\pi^0 X$,  
with the intermediate meson $R$ give a sizable contribution to the inclusive 
$\pi^0$ production cross section. In addition to the direct process a 
reasonable estimation for 
the inclusive cross section $\gamma\gamma\rightarrow\pi^0 X$ should contain 
the sum of dominant contributions from the intermediate meson resonances which 
contain at least one $\pi^0$ meson in the final state. 
We call this approach a Dominant Resonance Production (DRP) model.

The exclusive differential cross section $d\sigma_R/dy$ to produce 
a single meson $R$ with mass $M_R$ is given by Eq.~(\ref{difoneres}). 
In the DRP approximation the inclusive cross section 
of the $\pi^0$  production through the decay of intermediate resonances is
\begin{equation} 
{d\sigma_{incl} \over dy}(\pi^0)=\sum\limits_{R,k}{d\sigma_R \over dy}~B_R^{(k)}
(R\rightarrow \pi^0)~n_R^{(k)}(\pi^0), 
\label{difincl}
\end{equation}   
\noindent 
where $B_R^{(k)}(R\rightarrow \pi^0)$ is the branching ratio for decay 
of the resonance $R$ to the channel $k$ which contains at least one $\pi^0$. 
$n_R^{(k)}(\pi^0)$ is the number of $\pi^0$'s  in the corresponding channel 
$k$. 
If the first step decay products contain another resonance, a similar 
expression may be written in turn for the second step decays 
producing $\pi^0$. Introducing the value 
\begin{equation} 
B^{out}_{R}= \sum\limits_{k}~B_R^{(k)}(R\rightarrow \pi^0)~n_R^{(k)}(\pi^0)
\end{equation}   
\noindent one can rewrite Eq.~(\ref{difincl}) in the following way:
\begin{equation} 
{d\sigma_{incl} \over dy}(\pi^0)=\sum\limits_{R}{d\sigma_R \over dy}~B^{out}_{R}.
\label{difinclshort}
\end{equation}   

Particle Data information relevant to the calculation of 
$\sigma_{incl}$ is given in Tab.~\ref{tab1}. We selected the resonances 
with relatively large widths $\Gamma_{M_R\rightarrow \gamma\gamma}$.  
The values of $\Gamma_{tot}$ and 
$B^{in}_{R}=\Gamma_{M_R\rightarrow \gamma\gamma}/ \Gamma_{tot}$ from 
review~\cite{PartDat} which are necessary to calculate 
$\Gamma_{M_R\rightarrow \gamma\gamma}$ are given for completeness.  
On the contrary to Tabs.~\ref{tab2}--\ref{tab4}, the decay probabilities 
and meson widths in Tab.~\ref{tab1} were taken according to the most 
recent version of Review of Particle 
Physics~\cite{PartDat}. We assumed that the decays 
$f_0(980)\rightarrow\pi\pi$  and
$f_0(1370)\rightarrow\pi\pi$ take place with 100\% 
probability since there is no quantitative information in Ref.~\cite{PartDat}
on other channels. Isospin conservation relations provide the probability 
of $1/3$ for $\pi^0\pi^0$ charge state in such $\pi\pi$ channels.   

The integrated inclusive cross section of $\pi^0$ production in 
PbPb collisions 
($R_1=R_2=7.75$ fm) at LHC energies is found to be 
$\sigma_{incl}(\pi^0)= 106$~mb 
in DRP model, while the exclusive one is only $36.25$~mb. 
As one can see, the contribution to the $\pi^0$ production via intermediate 
resonances turns out to be essential. The corresponding rapidity distributions, 
$d\sigma/dy$, will be discussed in Sec.~\ref{rapid}.

\subsection{Pion production in $\gamma A$ collisions} 
 
The meson production induced by equivalent photons in electromagnetic 
collisions of ultrarelativistic heavy ions (Fig.~\ref{NNproc2}) is a 
poorly explored phenomenon. To the best of our knowledge, the first 
calculations of the  
total rate of pion production in electromagnetic collisions were made 
in Ref.~\cite{Bertulani86}. The first experimental evidence of  
electromagnetic dissociation accompanied by the pion production was found in 
Ref.~\cite{Singh} for 200 AGeV $^{16}{\rm O}$ ions. Due to a small number 
of pion production events detected in nuclear emulsion, the absolute cross 
section for such dissociation channel was not determined. To date only the 
existence of pion production in $\gamma A$ collisions is demonstrated by 
the experiment~\cite{Singh}.  
 
Let us consider the absorption by a nucleus of an equivalent photon 
leading to the pion production (Fig.~\ref{NNproc2}). The nucleus may absorb 
one or more virtual photons during a collision. We follow Llope and  
Braun-Munzinger~\cite{Llope} in description of such multiple absorption  
processes.  The double differential cross section of the pion 
production via the single photon absorption may be written first in the 
projectile rest frame. In such a frame the nucleus is at rest prior to  
absorption:  
\begin{equation} 
\frac{d^2\tilde\sigma^{(1)}}{p_tdp_tdy}= 
\int\limits_{\omega_{min}}^{\infty} 
\frac{d\omega_1}{\omega_1} N^{(1)}(\omega_1) 
\sigma_{A_2}(\omega_1)\frac{d^2W}{p_tdp_tdy}(\omega_1), 
\label{ga1} 
\end{equation} 
where $\sigma_{A_2}$ is the total photoabsorption cross section 
for the nucleus $A_2$ and $d^2W/p_tdp_tdy$ is the double differential  
distribution of pions produced by a photon with energy $\omega_1$. 
Spectral function $N^{(1)}$ is given by the following expression: 
\begin{equation} 
N^{(1)}(\omega_1)=2\pi\int\limits_{b_{min}}^{\infty} bdb  
e^{-m(b)} N(\omega_1,b), 
\label{ga2} 
\end{equation} 
where $N(\omega_1,b)$ is defined by Eq.~(\ref{aspectr}) and $m(b)$ is the 
mean number of photons absorbed by the nucleus $A_2$ in a collision at 
impact parameter $b$: 
\begin{equation} 
m(b)= 
\int\limits_{\omega_{min}}^{\infty}N(\omega ,b) 
\sigma_{A_2}(\omega)\frac{d\omega}{\omega} . 
\label{ga3} 
\end{equation} 
Analogously, for the second order process with a pair of photons absorbed by 
the nucleus $A_2$ the double differential cross section is: 
\begin{equation} 
\frac{d^2\tilde\sigma^{(2)}}{p_tdp_tdy}= 
\int\limits_{\omega_{min}}^{\infty}\int\limits_{\omega_{min}}^{\infty} 
\frac{d\omega_1}{\omega_1}\frac{d\omega_2}{\omega_2}  
N^{(2)}(\omega_1,\omega_2)\sigma_{A_2}(\omega_1)\sigma_{A_2}(\omega_2) 
\frac{d^2W(\omega_1)}{p_tdp_tdy}\frac{d^2W(\omega_2)}{p_tdp_tdy}, 
\label{ga4} 
\end{equation} 
with the corresponding double photon spectral function: 
\begin{equation} 
N^{(2)}(\omega_1,\omega_2)=\pi\int\limits_{b_{min}}^{\infty} bdb  
e^{-m(b)} N(\omega_1,b) N(\omega_2,b). 
\label{ga5}  
\end{equation} 
 
In the above expressions $b_{min}$ is the minimal value of the impact 
parameter which corresponds to the onset of nuclear interaction. 
In pion production calculations the 
integration over the photon energy starts from $\omega_{min}\approx 140$ MeV 
which corresponds to the pion emission threshold.
Finally, with the corresponding 
Lorentz boost from the nucleus rest frame to the laboratory system one can 
obtain the double differential distribution $d^2\sigma/{p_tdp_tdy}$ for 
produced pions suitable for measurements in experiments.
Further details of our approach may be found in  
Refs.~\cite{Pshenichnov,Pshenichnov2}. 
 
As it was shown in Ref.~\cite{Pshenichnov2}, the contribution 
to pion production from the double photon absorption is less then 10\%  
for heavy ions with  $\gamma\gg 100$. Since the contribution  
from the third and fourth order processes are expected to be even lower, 
these processes can be safely disregarded for the ultrarelativistic energies. 
 
Because of the coherent action of all the charges in the nucleus, 
the virtuality of the emitted photon, $Q^2=-q^2$, is restricted.  
Such photons are almost real, $Q^2\leq 1/R^2$, where $R$ is 
the nuclear radius.
Therefore, photonuclear data obtained in experiments with monoenergetic 
photons may be used, in principle, as an input for Weizs\"{a}cker-Williams 
calculations of pion production in  $\gamma A$ collisions. Since the  
spectrum of equivalent photons covers a very wide range of the photon  
energies, one needs the double differential distribution  
$d^2W(\omega)/p_tdp_tdy$ for photon energies $\omega$ starting from the  
pion emission threshold and up to several tens or even hundreds of GeV. 
 
In the region of interest, i.e. at $\omega\geq 140$ MeV, the photon de  
Broglie wavelength is comparable or even smaller than the nucleon radius. 
A photon interacts, mainly, with individual intranuclear nucleons.  
Experimental data on single pion photoproduction on the nucleon are 
accumulated in compilations of Refs.~\cite{Ukai,Alekhin}. 
The latter compilation contains also the experimental data on  
photoproduction of baryon, $B^\star$, and meson $M^\star$ resonances: 
$\gamma N\rightarrow\pi B^\star$ and $\gamma N\rightarrow\pi M^\star$ 
as well as on some channels of multiple pion production: 
$\gamma N\rightarrow i\pi N$, $2\leq i\leq 8$. 
Due to a long mean free path an equivalent photon may be 
absorbed deeply in the nuclear interior. 
The mesons produced in a photonucleon reaction interact with other intranuclear 
nucleons inducing different reactions in the nuclear medium.  
In other words, the process shown in Fig.~\ref{NNproc2} is followed by  
the final state interaction of produced hadrons   
with the residual nucleus $A_2$. Beside the pions several nucleons may be 
emitted and the residual nucleus may receive a high excitation energy.    
 
Since the data on pion photoproduction on nuclei exist only for limited 
domains of $\omega$ and pion kinematical variables, a theoretical  
model should be used in the cases when the data are not available. 
A suitable tool for describing such 
multi-step photonuclear reactions is the Intranuclear Cascade Model, which is 
well-known for many years~\cite{Barashenkov}. 

As shown in Ref.~\cite{Iljinov}, the extended Intranuclear Cascade  
Model of photonuclear reactions describes reasonably well available 
data on meson production and nucleon emission 
obtained in last two decades with  
intermediate energy quasi-monochromatic photons. With this in mind one can 
use the INC model for the Weizs\"{a}cker-Williams calculations, as it was 
demonstrated in Refs.~\cite{Pshenichnov,Pshenichnov2}. Formally we use 1 TeV 
as the upper limit for the equivalent photon energy in the $\gamma A$ process. 
Our model was not initially designed for such high energies 
when events with a very high hadron multiplicity become possible and many other
channels, like baryon-antibaryon photoproduction, may be open. 
Nonetheless one can safely use the model to investigate the $\gamma A$
processes with low multiplicity which mimic processes from 
$\gamma\gamma$ fusion. The former processes take place mainly at 
$\omega <10$ GeV, where our model has been verified in detail.

\subsubsection{Simulation of $\gamma N$ interaction} 
 
The channels of the hadron photoproduction on a nucleon which are taken into 
account in the model are listed in Tab.~\ref{gat1}.  
To describe the two-body photoproduction channels we have basically used the  
Monte Carlo event generator of Corvisiero et al.~\cite{Corvisiero}. 
The contribution from the 
two-body channel $\gamma N\rightarrow\pi N$ dominates up to  
$\omega\approx 0.5$ GeV. Approximations of the total and  
differential cross sections for such channels based on the model 
of Walker and Metcalf~\cite{Metcalf} were used up to 2 GeV.  
The excitation of six baryon resonances was taken into account. 
The contributions from $\Delta(1232)$, $N^\star (1520)$ and $N^\star (1680)$ 
are the most important. The presence of these resonances explains a well-known 
resonant structure in the total $\gamma N$ cross section at 
$\omega\leq 1.2$ GeV.  
 
The channels $\gamma N\rightarrow 2\pi N$ and $\gamma N\rightarrow 3\pi N$ 
play a major role at $0.5\leq \omega \leq 2$ GeV. These channels include also 
the resonant contributions from $\pi\Delta$, $\eta N$, $\rho N$ and 
$\omega N$ channels (Tab.~\ref{gat1}). Although the presence of these  
contributions in the total $\gamma N$ cross sections is difficult to trace, 
the angular distributions of these channels have some specific features. 
Several examples may be given. The $\gamma p\rightarrow\eta p$ channel 
has three important contributions from $S_{11}(1535)$, $S_{11}(1700)$ and 
$P_{11}(1750)$ states. Due to the dominance of $S_{11}(1535)$ the angular 
distribution of the process is not far from isotropy. On the contrary, 
$\gamma p\rightarrow \rho^0 p$ process has a prominent forward peak,  
since the non-resonant diffractive contribution dominates.  
 
Many channels in the multi-pion reactions,
$\gamma N\rightarrow i\pi N$, ($2\leq i\leq 8$), were not  
suitable for measurements in spark, bubble or streamer chamber experiments  
due to the presence of several neutral particles in the final state. 
However, one can reconstruct the integral cross sections 
of undetectable channels by applying isotopic relations to the  
measured cross sections of channels with charged particles which 
can be found in compilation of Ref.~\cite{Alekhin}. 
A phenomenological statistical approach for the exclusive description of the  
elementary $\gamma N$ interaction was applied to multiple pion production  
channels in Ref.~\cite{Iljinov}. There an isospin statistical model was 
used to connect unknown cross sections with measured ones.  
Other details of the method used for simulating the $\gamma N$ interaction  
may be found elsewhere, see Ref.~\cite{Iljinov}.   
 
Since a huge number of multiple pion production channels is open at  
$\omega > 2 $ GeV, the statistical description may be the only way to  
estimate the cross sections of such channels.  
Recently another kind of statistical assumptions was used  
in Monte Carlo event generator Sophia for simulating  photohadronic 
processes in astrophysics~\cite{Mucke}.

\subsubsection{Secondary processes} 

According to our model the fast hadrons produced in a primary $\gamma N$ 
interaction initiate a cascade of successive quasi-free hadron-nucleon 
collisions inside the nucleus. The following  
elementary processes were taken into account: 
$$ 
    \pi N \rightarrow \pi N\, , \,  
     \pi (NN) \rightarrow NN\, , \,  
     \pi N \rightarrow \pi \pi N\, ; 
     \, \pi N \rightarrow (i+1) \pi N \, \,  , $$ 
$$ 
   NN \rightarrow NN\, , \,                  
    NN \rightarrow \pi NN\, , \, 
    NN \rightarrow i \pi NN \, \, , (i \geq 2)\, ; 
$$ 
$$ \eta N \rightarrow \eta N\, , \,              
    \eta N \rightleftharpoons \pi N\, , \,     
    \eta N \rightarrow \pi \pi N\, , \,  
    \eta (NN) \rightarrow NN\, ,  
    \eta (NN) \rightarrow \pi NN\, , \,  $$ 
$$ \omega N \rightarrow \omega N\, , \,          
    \omega N   \rightleftharpoons  \pi N\, ,    
    \omega N \rightarrow \pi \pi  N \, , \,  
    \omega (NN) \rightarrow NN\, , \,  
   \omega (NN) \rightarrow \pi NN\, .
$$ 
The empirical approximations for the measured integral and differential 
cross sections  
of $NN$ and $\pi N$ interactions as well as the phenomenological estimations 
for the total and partial cross sections of $\eta N$ and 
$\omega N$ interactions were used in the calculation~\cite{Iljinov}.
      
The described Monte Carlo model is implemented into  the RELDIS  
code~\cite{Pshenichnov2} which is especially designed for calculating the  
electromagnetic dissociation processes in ultrarelativistic heavy ion  
collisions.

\subsubsection{Characteristics of produced pions} 
 
The average number of pions, $\langle n_\pi\rangle $, produced in 
a $\gamma A$ process is quite small.  
The values of $\langle n_\pi \rangle$ for each pion charge are 
given in Tab.~\ref{gat2} for AuAu and PbPb collisions at RHIC and LHC  
energies, respectively.  
Even with the inclusion of multiple pion production channels 
($\gamma N\rightarrow i\pi N$, $2\leq i\leq 8$),  
the average numbers of pions of each charge are quite low ($\approx 1$).  
This can be explained by two reasons.  First, in a great part of the 
electromagnetic dissociation events no pions are produced at all, due to 
the dominance of soft 
photons ($\omega \leq 140$ MeV) in the equivalent photon spectrum, 
Eqs.~(\ref{ga2}),(\ref{ga5}). Second, the most probable processes  
of pion production are $\gamma N\rightarrow\pi N$ and  
$\gamma N\rightarrow\pi \Delta$, which produce one or two pions only.  
  
As one can see from Tab.~\ref{gat2}, in the $\gamma A$ process 
the neutral pions are produced more abundantly as compared with $\pi^+$ and 
$\pi^-$ mesons. A difference is noticeable also in the differential 
distributions, $d\sigma/dy$, shown in Fig.~\ref{gaf1} for each pion charge. 
The main deviations in the pion yields appear at rapidities close to the 
beam rapidity. This fact has a simple explanation. 
As already mentioned above, the process $\gamma N\rightarrow\pi N$ 
dominates in the pion production by equivalent photons
and corresponding pions are close to the beam rapidity. 
This reaction can proceed on a proton: $\gamma p\rightarrow\pi^+ n$, 
$\gamma p\rightarrow\pi^0 p$ and a neutron: $\gamma n\rightarrow\pi^- p$, 
$\gamma n\rightarrow\pi^0 n$. Thus, 
single neutral pions can be produced both on the proton and the neutron, 
while $\pi^+$ on the neutron and $\pi^-$ on the proton only. The total  
cross sections of these four channels are close to each other.  
This gives the rate of $\pi^0$ production approximately twice as large as the  
rate of $\pi^+$ or $\pi^-$ for a light nucleus with the equal numbers 
of protons and neutrons. This feature of the single pion photoproduction is 
confirmed by the measurements of inclusive cross sections and the 
calculations made for carbon nucleus (see Fig. 15 of Ref.~\cite{Iljinov}).   
 
Since a neutron excess exists in heavy nuclei like Pb or Au, in this case
the $\pi^-$ production will be enhanced compared to $\pi^+$, as it is  
shown in Fig.~\ref{gaf1}.  The final state interaction affects the absolute  
yields of pions of different charges, but does not essentially change the  
ratio between them.  
 
The detection of pions produced close to the beam rapidity is a  
complicated experimental task. Special zero-degree detectors with  
a proper identification of particle mass and charge are necessary for this  
purpose. However if a forward detector located after a steering magnet 
is suitable for determination of charges of nuclear fragments, one may 
exploit it to detect the pion emission process indirectly.
If $\pi^-$ is produced in the reaction  
$\gamma n\rightarrow\pi^- p$, it may be emitted while the recoil proton 
may be captured by the residual nucleus. The deexcitation of such 
nucleus may take place mainly via neutron emission. It means that the  
initial charge of the nucleus will be increased by one unit. Charge-to-mass 
ratio will be changed for such ion and it will be separated from the 
beam. Depending on the value of heavy-ion energy RELDIS code predicts the 
cross section of such electromagnetic ``charge pick-up'' channels at the level 
of 10--100 mb. This process competes with the nucleon pick-up 
process via strong interaction.  Nevertheless as it is known from  
Ref.~\cite{KLM}, the cross section  
of the latter process drops noticeably with increasing beam energy.  
On the contrary, the cross section of electromagnetic charge pick-up 
increases gradually with the beam energy and may essentially 
exceed the value of cross section for the nucleon pick-up due to the 
strong interaction.

The photonuclear reactions $(\gamma,\pi^-xn)$, $x=0-9$ induced by 
bremsstrahlung photons were discovered many years ago. We refer the 
reader to a recent paper where such
reactions were studied by radiochemical methods~\cite{Sakamoto}. It would be 
interesting  to study the same type of reactions induced by 
equivalent photons in ultrarelativistic heavy ion collisions.  
               
As mentioned above, the main part of pions is produced close to the 
beam rapidity. A small fraction of pions is produced by a very-high 
energy photons with $\omega \gg 10$ GeV.  Some of the pions produced in such  
interactions may receive the momentum high enough to populate the  
central rapidity region. As shown in Fig.~\ref{gaf1}, there are  
pions with $|y|<5$ which may be confused with those from $\gamma\gamma$  
process. Therefore some selection criteria are necessary for 
$\gamma\gamma$ processes (see Sec.~\ref{rapid}).

\section{Very peripheral nuclear collisions} 
 
Very peripheral (grazing) nuclear collisions with the participation 
of strong nuclear forces can be misidentified as the electromagnetic 
interaction events. Both of these types of interactions contribute to the 
events with low multiplicity of particles. 
To study the properties of strong interaction events a Monte Carlo 
event generator FRITIOF, version 7.1~\cite{FRITIOF} is used. 

\subsection{Basic processes considered by FRITIOF model}
 
Let us recall the main statements of the extended FRITIOF 
model~\cite{FRITIOF} which
is valid up to the TeV region. FRITIOF is a Monte Carlo model for 
hadron-hadron, hadron-nucleus and nucleus-nucleus 
collisions. The basic idea of the model is in a simple picture that 
a hadron behaves like a relativistic string with a confined color field. 
This field is equivalent to that of a chain of dipoles lined up along the 
axis line. The dipole links act as partons. During the soft interaction 
many small transverse momenta are exchanged between the
dipole links and two longitudinally excited string states result from the 
collisions. A disturbance of the color field will in general initialise 
the gluonic radiation according to QCD. The final state particles are 
obtained by fragmentating the string states like in the $e^+e^-$ 
annihilation.

The large $p_t$ process can be treated by using QCD directly. 
The hard interaction effects, which are considered as the Rutherford parton 
scattering, become important in the TeV range of c.m. energies and at 
large  $p_t > 1 $ GeV/c. 
The results of the model are in good agreement with experimental hadron-hadron 
data up to the highest energies currently available.

Nucleus-nucleus collisions are regarded in the FRITIOF model as incoherent 
collisions between nucleons of colliding nuclei. 
FRITIOF does not take into account collective (coherence) effects when 
two nuclei interact as a whole. Thus a nucleon from the 
projectile interacts independently with the encountered target nucleons 
as it passes through the nucleus. Each of these 
subcollisions can be treated as a usual hadron-hadron collision. 
On the time scale of the collision process, the exited nucleon does not 
fragment inside the nucleus, so there are no intranuclear cascades. 
This assumption is reasonable at high energy since the time scale 
associated with fragmentation is much longer than the flight 
time of excited nucleons through the nucleus.

\subsection{Nuclear density distributions and total cross section in PbPb 
collisions}

One of important aspects of the collision is the nuclear geometry.  
It is assumed that the projectile passes through the 
target nucleus on a straight line trajectory. The nucleons are distributed 
inside the nucleus according to the nuclear density distribution $\rho(r)$.
Wood-Saxon density function for heavy nuclei was used in our calculations:
\begin{equation}
\rho(r)={\rho_o\over1+exp({r-r_oA^{1/3}\over C})},
\end{equation}
where $r_o=1.16(1-1.16A^{-2/3})$ is the radial scale parameter, 
$C$ is the diffuseness parameter,
it is slightly $A$-dependent and its values used in the program ranging from 
0.47 to 0.55~fm, and $\rho_o$ is a normalisation constant. 
Then for lead nucleus the radius defined at the half of the normal 
nuclear density is equal to $R=r_oA^{1/3}=6.63 $~fm, 
while the diffuseness parameter used by FRITIOF is equal to $C=0.545$~fm.     

The total cross section of strong lead-lead interaction at LHC collider
was estimated to be 7164 mb, according 
to a simple geometrical formula used in Ref.~\cite{HePrice}. 
This estimation does 
not take into account the energy dependence of the total nucleus-nucleus
cross section due to the increase of the total nucleon-nucleon cross section.
Basing on extrapolations of the measured $pp$ cross sections to LHC energies
one can obtain a higher value of about ~10 barn, see Ref.~\cite{Lokhtin}.
For this value of the total nucleus-nucleus cross section  one has to 
rescale the plots given in the present paper by ~30\%  
for grazing nuclear collisions.

\subsection{Impact parameter distribution in nuclear collisions} 

Special efforts should be undertaken to find  proper ranges of the impact
parameter $b$ which divide nucleus-nucleus collisions 
into different categories. These categories were listed in the introduction,
most important for us are 
very peripheral nuclear collisions and electromagnetic interactions, 
subsequently subdivided into $\gamma\gamma$ and $\gamma A$ events.  

The uncertainties associated with the diffuseness of the nuclear density
distribution affect grately the resulting ranges. We used the FRITIOF model 
to define these ranges. The impact parameter distribution of events simulated 
by the FRITIOF code is given in Fig.~\ref{impact0}.   
The plotted quantity $dN_{AA}(b)/db/N_{AA}$ is a relative number of nuclear 
collisions with given $b$.
As it is shown in the figure, the events with a low multiplicity of pions
correspond to collisions with $12 \leq b \leq 20$~fm. It is the domain
of $b$ which defines the grazing nuclear collisions with participation 
of strong nuclear forces. It should be stressed however, that this 
definition is essentially model dependent and numerical results may be 
different if one uses another code with different treatment of the diffuse 
boundary of the nucleus and another values of the elementary $NN$ 
cross sections.  

We see that the value which defines the onset of electromagnetic collision
for PbPb ions, $b_{min}=15.5$~fm, used in the paper~\cite{Pshenichnov} 
does not contradict the results of the FRITIOF calculations given in 
Fig.~\ref{impact0}.  Following our previous 
papers~\cite{Pshenichnov,Pshenichnov2},
for consistency we define the end of nuclear overlap in lead-lead 
collisions at $b_{min}= 15.5$ fm.  This choice is also supported by
the calculation of Ref.~\cite{Pajares} where the average number of interacting 
nucleons, $\langle N_{in}\rangle$, was evaluated  as a function of $b$ 
for the case of ${\rm PbPb}$ collisions.
As it was found~\cite{Pajares}, $\langle N_{in}\rangle\approx 1$ at 
$b=15.5$ fm. According to the Wood Saxon density function for ${\rm Pb}$ 
nucleus it corresponds to the overlap of those boundary regions of 
the colliding nuclei which have the nuclear densities below 0.1 of 
the value at the center of the nucleus. 
The total number of such events in the FRITIOF simulation turns out to be 
24\%.
 
Fig.~\ref{N_n_tot} demonstrates the distribution of pion multiplicities 
for peripheral PbPb collisions ($b>15.5$~fm) at LHC energies
simulated by the FRITIOF code. The most probable is
the production of four pions and the distribution is very broad.
This conclusion is valid for all the charge states of pions.  
The average pion multiplicities are given in 
Tab.~\ref{gat2}, where one can see that the multiplicities of pions 
produced in the above-defined grazing $AA$ collisions are generally 
greater than those for $\gamma A$ process.

\section{Comparison of different mechanisms of pion production}\label{rapid} 
   
Since most of the detectors used thus 
far in heavy-ion experiments have 
very restricted acceptance on rapidity, it is important to
investigate the rapidity dependence of pion production cross section.
Rapidity distributions of neutral pions produced in 
peripheral PbPb collisions at LHC energies are shown in 
Fig.~\ref{incl0}. 

Let us consider first the distributions without any additional
selection criteria. The distributions from $\gamma\gamma$ fusion and grazing 
nuclear collisions calculated  by the FRITIOF code with 
$b\geq 15.5$ fm are very similar in shape.
The distributions of $d\sigma/dy$ for single and inclusive $\pi^0$
production in $\gamma\gamma$ fusion also have maximuma at $y=0$. 
But even in this region it is much lower then the contribution from 
grazing nuclear collisions. 
On the contrary, the rapidity distribution for $\gamma A$ production 
has peaks at the beam and target rapidities. Nevertheless its contribution 
at midrapidity is comparable to the one from $\gamma\gamma$ fusion. 

This above-described picture is unfavourable for experimental detection
of pions from $\gamma\gamma$ collisions. Nonetheless
one can improve the detection conditions by using appropriate 
transverse momentum cuts. Indeed the single mesons from 
$\gamma\gamma$ fusion have a very small transverse momenta. 
Selecting events with low values of $p_t$ one can reject contributions from
other processes. Following this way two different $p_t$--cut 
procedures may be proposed.  

According to the first procedure (b-criterion) one should select 
the events with a small total transverse 
momentum of the meson system $|\vec {p_t}^{(sum)}|\leq 75$~MeV/c. 
The second selection procedure (c-criterion) imposes more 
severe restrictions on the transverse momenta of the produced  pions 
demanding each of them (including $\pi^0$, $\pi^+$, and $\pi^-$) to be small, 
$|\vec {p_t}|\leq 75$ MeV/c, 
in this case. As one can see, b-criterion is equivalent to 
c-criterion in the case of single pion production.
    
Let us investigate now to what extent the rapidity distributions are 
affected by 
$p_t$ cuts. The distributions obtained according to b- and c-criteria
are also given in Fig.~\ref{incl0}. One can see that the b-criterion 
reduces the contribution from grazing nuclear collisions by three orders 
of magnitude and
give some benefits for detection of $\gamma\gamma$ events.
If this reduction is not sufficient one can use c-criterion which is more efficient. 
Further suppression of pions from grazing nuclear collisions may be obtained 
in the region $-4<y<4$. With this suppression even exclusive 
$\gamma\gamma\rightarrow\pi^0$ channel may be clearly distinguished.

The difference in b- and c-criteria applied to $\gamma A$ events calculated by 
RELDIS code is less prominent. This is explained by the fact that
single pion production dominates in $\gamma A$ collisions at $\omega\geq 140$ MeV
while b- and c-criteria are equivalent for this case. 
Both of the procedures may be recommended to reduce the contribution from
$\gamma A$ process as it is shown in Fig.~\ref{incl0}. 

It should be noticed that b-criterion does not affect the inclusive
distribution for $\gamma\gamma$ fusion, while c-criterion suppresses the main
part of events of $\pi^0$ production due to heavy meson decays. This may be
useful for extraction of the exclusive $\gamma\gamma\rightarrow\pi^0$ process.

\section{Conclusions} 
 
Only several hadrons are produced on average in very peripheral 
collisions of ultrarelativistic heavy ions. Because of this feature such 
collisions can be considered as non-violent events. An  
exact determination of the impact parameter in a collision event is beyond  
present experimental techniques. However, as it is shown in the 
present paper, selecting the events with very low multiplicity of produced 
mesons one can approximately identify a domain of large impact parameters 
where $\gamma\gamma$ fusion, $\gamma A$ or grazing nuclear $AA$ collisions 
takes place. 
 
Calculations show that each of the mechanisms has specific distribution of  
produced pions on transverse momentum, $p_t$, and rapidity, $y$. One can 
use these 
features for adjusting detectors to different regions of $p_t$ and $y$ 
with possibility to disentangle pions produced by different mechanisms. 
One can enhance 
the signal to background ratio for $\gamma\gamma$ fusion by selecting 
events with low $p_t$ by means of two procedures considered in the paper. 
Our preliminary results confirm the importance of $p_t$ cut proposed in 
Ref.~\cite{CMS}. This procedure rejects the background from 
$\gamma A$ and grazing $AA$ collisions by several orders 
of magnitude. 
Moreover using $p_t$ cuts one can avoid a pessimistic conclusion made in 
Ref.~\cite{Engel} that the $\gamma\gamma$ fusion is indistinguishable from
other process since the rapidity distributions for different 
processes are very similar to each other.    

\section{Acknowledgments}
 
L.I.S. and A.P.K. are grateful to Dr. Kai Hencken for useful discussions
and a possibility to use his code for comparison.      
I.A.P. is indebted to INTAS for the financial support from Young Scientists 
Fellowship 98-86 and thanks the Niels Bohr Institute for the warm
hospitality.  
The work is supported in part by the Universities of Russia Basic  
Research Fund, grant 5347, RFBR-DFG grant 99-02-04011 and by the 
Humboldt Foundation, Germany.

 

\clearpage 

\begin{table} 
\caption{ 
Exclusive cross sections of single meson production in $\gamma \gamma$ fusion
process in ultrarelativistic UU and PbPb collisions at colliders. 
The values were  
calculated for comparison with Ref.~\protect\cite{BaurFilho}. 
The masses and two-photon 
decay widths for pseudoscalar mesons were used according to 
Ref.~\protect\cite{BaurFilho}.
} 
\begin{center} 
\begin{tabular}{|c|c|c|c|c|c|c|} \hline 
\multicolumn{3}{|c|}{  } & \multicolumn{4}{c|}{cross section (mb)} \\ \hline  
 
Meson & $M_R$ & $\Gamma_{M_R\rightarrow \gamma\gamma}$   & 
\multicolumn{2}{c|}{Z=92 A=238 $\gamma=100$} & 
 \multicolumn{2}{c|}{Z=82 A=208 $\gamma=4000$} \\ \cline{4-7} 
           & (GeV)    & (keV)   & \cite{BaurFilho}  
                  & this work  & \cite{BaurFilho} & this work  \\ \hline 
$\pi^0$  & 0.135 & 0.009 & 7.9 & 9.55 & 49    &53.1 \\ 
$\eta$   & 0.549 & 1     & 2.9 & 3.61 & 43   & 45.8 \\ 
$\eta'$  & 0.958 & 5     & 1.1 & 1.48 & 30    & 31.9 \\ 
$\eta_c$ & 2.981 & 6.3  & $2.5\cdot 10^{-3}$ & $4\cdot 10^{-3}$ & 0.59 & 0.644 \\ 
$\eta_b$ & 9.366 & 0.41 & $7\cdot$ $10^{-9}$ & 3.36$\cdot$ $10^{-8}$ & 0.46   & 
$5\cdot 10^{-4}$ \\ 
\hline 
\end{tabular} 
\end{center} 
\label{tab2}  
\end{table} 

\begin{table} 
\caption{ 
Exclusive cross sections of single meson production in $\gamma \gamma$ fusion
process in ultrarelativistic PbPb collisions at LHC.
The values were calculated for comparison with Ref.~\protect\cite{BaurHen}. 
The masses and 
two-photon decay widths for $c\bar{c}$ and $b\bar{b}$ mesons 
were used according to Ref.~\protect\cite{BaurHen}.
}  
\begin{center} 
\begin{tabular}{|c|c|c|c|c|}  
\hline 
\multicolumn{3}{|c|}{  } & \multicolumn{2}{c|}{cross section (mb)} \\ \hline  
 
Meson & $M_R$ & $\Gamma_{M_R\rightarrow \gamma\gamma}$ & 
\multicolumn{2}{c|}{Z=82 A=208 $\gamma=2750$}\\  
\cline{4-5} 
           & (GeV)    & (keV)   & \cite{BaurHen}  & this work  \\ \hline 
$\eta'$ &0.958 & 4.2 & 22 & 20.68 \\ 
$\eta_c$ & 2.981 & 7.5 & 0.59 & 0.552 \\ 
$\chi_{0c}$ & 3.415 & 3.3 & 0.16 & 0.145 \\  
$\eta_b$ & 9.366 & 0.43 & 3.7$\cdot 10^{-4}$ & 3.46$\cdot 10^{-4}$ \\ 
$\eta_{0b}$ & 9.860 & 2.5$\cdot 10^{-2}$ & 1.8$\cdot 10^{-5}$ & 1.62$\cdot 10^{-5}$ \\ 
$\eta_{2b}$ & 9.913 & 6.7$\cdot 10^{-3}$ & 2.3$\cdot 10^{-5}$ & 2.13$\cdot 10^{-5}$ \\  
\hline 
\end{tabular} 
\end{center} 
\label{tab3}  
\end{table} 
 
\begin{table} 
\caption{
Exclusive cross sections of single meson production in $\gamma \gamma$ fusion
process in ultrarelativistic AuAu and PbPb collisions at RHIC and LHC.
The values were calculated for comparison with Ref.~\protect\cite{Krauss}. 
The masses and 
two-photon decay widths for different scalar and pseudoscalar mesons were used 
according to Ref.~\protect\cite{Krauss}.
} 
\begin{center} 
\begin{tabular}{|c|c|c|c|c|c|c|} \hline 
\multicolumn{3}{|c|}{  } & \multicolumn{4}{c|}{cross section (mb)} \\  
\hline  
Meson   & $M_R$ & $\Gamma_{M_R\rightarrow \gamma\gamma}$  & 
\multicolumn{2}{c|}{Z=79 $\gamma=108$} &  
\multicolumn{2}{c|}{Z=82 $\gamma=2750$}\\  
\cline{4-7} 
           & (GeV)    & (keV)   & this work   & \cite{Krauss}   
           & this work   & \cite{Krauss} \\ \hline 
$\pi^0$ & 0.135 & 0.0077  & 4.744 & 5.721 & 37.701 & 42.937\\ 
$\eta $ & 0.547 & 0.51    & 1.127 & 1.285 & 18.744 & 19.897 \\ 
$\eta'$ & 0.958 & 4.5     & 0.826 & 0.989  & 22.152 & 24.780 \\ 
$f_0(975)$ & 0.974 & 0.25 & 0.042   & 0.0909 & 1.159 & 2.496 \\ 
$f_0(1250)$ & 1.25 & 3.4  & 0.177 & 0.332   & 6.360 & 11.728 \\ 
$f_2$ & 1.275 & 3.19      & 0.757 & 0.679   & 27.751 & 24.674 \\ 
$a_2 $ & 1.318 & 1.14     & 0.230 & 0.252   & 8.784 & 9.542 \\ 
$\pi_2 $ & 1.670 & 1.41   & 0.087 & 0.104   & 4.547 & 5.188 \\ 
$f_4 $ & 2.05 & 1.4       & 0.053 & 0.0221   & 3.795 & 1.605  \\ 
$\eta_c $ & 2.98 & 6.3    & $3.18\cdot 10^{-3}$ & $3.66\cdot 10^{-3}$  & 0.464 & 0.555 \\ 
$\chi_{0c}$ & 3.42 & 5.6   & $1.21\cdot 10^{-3}$ & $1.36\cdot 10^{-3}$ & 0.243 & 0.290 \\ 
$\eta_b$ & 9.37 & 0.4 & $3.15\cdot 10^{-8}$ & 2$\cdot 10^{-8}$ & 
$3.22\cdot 10^{-4}$ &  $4.06\cdot 10^{-4}$ \\ \hline 
\end{tabular} 
\end{center} 
\label{tab4}  
\end{table} 

\begin{table} 
\caption{ 
Resonances, their decay modes and branching ratios (given in brackets) which were taken 
into account in inclusive cross section calculation in DRP model.} 
\vspace{0.3cm} 
\small{ 
\begin{tabular}{|c|c|c|c|c|c|c|} \hline 
 & & & & & & \\
Channel & $M_R$ & $\Gamma_{tot} $ & $B^{in}_{R}$ & 1st decay step 
& 2nd decay step & $B^{out}_{R}$ \\ 
& (GeV) & (GeV) & & & & \\ 
\hline 
$\gamma\gamma\rightarrow\pi^0$ & 0.135 & 7.0$\cdot 10^{-9}$ & 1. & $\pi^0 $ & & 1 \\ 
\hline
$\gamma\gamma\rightarrow\eta$ & 0.547 & 1.18$\cdot 10^{-6}$ & 0.388 & 
$\eta\rightarrow 3\pi^0$ (0.32) & &1.19\\ 
\cline{5-5} & & & & $\eta \rightarrow \pi^+\pi^- \pi^0$ (0.23) & & \\ 
\hline 
$\gamma\gamma\rightarrow\eta'$ & 0.958 & 2.03$\cdot 10^{-4}$ & 0.021 & 
$\eta'\rightarrow \pi^0\pi^0\eta$ (0.21) &$\eta \rightarrow 3\pi^0$ (0.32) & 1.18 \\ 
\cline{6-6} & & & & & $\eta \rightarrow \pi^+\pi^-\pi^0$ (0.23)  & \\ 
\cline{5-6} & & & &  $\eta' \rightarrow \pi^+\pi^-\eta $ (0.44) &
 $\eta \rightarrow 3\pi^0$ (0.32) &  \\ 
\cline{6-6} & & & & & $\eta \rightarrow \pi^+\pi^-\pi^0$ (0.23) &  \\ 
\hline 
$\gamma\gamma \rightarrow f_0$ & 0.980  & 0.07   & 1.19$\cdot10^{-5}$ & 
$f_0\rightarrow\pi^0\pi^0$ (0.33)  & & 0.66 \\ 
\hline 
$\gamma\gamma \rightarrow f_2$ & 1.275  & 0.185 & 1.32$\cdot10^{-5}$ & 
$f_2 \rightarrow \pi^0\pi^0$ (0.28)  & & 0.70 \\ 
\cline{5-6} 
 & & & & $f_2\rightarrow \pi^+\pi^-2\pi^0$ (0.07)  & &  \\ 
\hline 
$\gamma\gamma\rightarrow a_2$ & 1.318  & 0.107   & 9.4$\cdot 10^{-6}$ & 
$a_2\rightarrow\eta\pi^0 $(0.15)  &$\eta\rightarrow 3\pi^0$ (0.32) & 0.33  \\ 
\cline{6-6} & & & & & $\eta\rightarrow\pi^+\pi^-\pi^0(0.23)$ &  \\ 
\hline 
$\gamma\gamma\rightarrow\pi_2$ & 1.670  & 0.258   & 5.6$\cdot 10^{-6}$ & 
$\pi_2 \rightarrow f_2\pi^0$ (0.56) &$f_2\rightarrow\pi^0\pi^0 $ (0.28) & 1.10 \\ 
\cline{6-6} & & & & &$f_2 \rightarrow \pi^+\pi^- 2\pi^0$ (0.07) &  \\ 
\cline{5-6} & & & &$\pi_2 \rightarrow f_0(1370)\pi^0$ (0.09) & 
$f_0(1370)\rightarrow \pi^0\pi^0(0.33)$ &  \\ 
\hline
\end{tabular} 
} 
\label{tab1} 
\end{table} 

\begin{centering} 
\begin{table} 
\caption{Channels of elementary $\gamma N$ interaction taken 
 into account in the INC model.} 
\vspace{0.3cm} 
\begin{tabular}{|c|c|} 
\hline 
               &         \\ 
$\gamma p$-interaction & $\gamma n$-interaction  \\ 
               &         \\ 
 \hline 
$\gamma  p \rightarrow \pi^+ n$   
&$\gamma  n \rightarrow \pi^- p$  \\  
$\gamma  p \rightarrow \pi^0 p$    
&$\gamma  n \rightarrow \pi^0 n$  \\  
               &         \\ 
$\gamma  p \rightarrow \pi^-  \Delta^{++}$  
&$\gamma  n \rightarrow \pi^-  \Delta^{+}$ \\ 
$\gamma  p \rightarrow \pi^0  \Delta^{+}$  
&$\gamma  n \rightarrow \pi^0  \Delta^{0}$ \\ 
$\gamma  p \rightarrow \pi^+  \Delta^{0}$  
&$\gamma  n \rightarrow \pi^+  \Delta^{-}$ \\ 
               &          \\ 
$\gamma  p \rightarrow \eta p$    
&$\gamma  n \rightarrow \eta n$  \\  
$\gamma  p \rightarrow \omega p$    
&$\gamma  n \rightarrow \omega n$  \\  
$\gamma  p \rightarrow \rho^0 p$    
&$\gamma  n \rightarrow \rho^0 n$  \\  
$\gamma  p \rightarrow \rho^+ n$    
&$\gamma  n \rightarrow \rho^- p$  \\ 
               &            \\ 
$\gamma  p \rightarrow \pi^+  \pi^-  p$  
&$\gamma  n \rightarrow \pi^+  \pi^- n $\\ 
$\gamma  p \rightarrow \pi^0  \pi^+  n$  
&$\gamma  n \rightarrow \pi^0  \pi^- p $\\ 
              &              \\ 
$\gamma  p \rightarrow \pi^0  \pi^0   \pi^0  p$  
&$\gamma  n \rightarrow \pi^0  \pi^0   \pi^0  n $\\ 
$\gamma  p \rightarrow \pi^+  \pi^-   \pi^0  p$  
&$\gamma  n \rightarrow \pi^+  \pi^-   \pi^0  n $\\ 
$\gamma  p \rightarrow \pi^+  \pi^0   \pi^0  n$  
&$\gamma  n \rightarrow \pi^-  \pi^0   \pi^0  p $\\ 
$\gamma  p \rightarrow \pi^+  \pi^+   \pi^-  n$  
&$\gamma  n \rightarrow \pi^+  \pi^-   \pi^-  p $\\ 
                      &           \\ 
$\gamma  p \rightarrow i \pi  N (4 \leq i \leq 8)$ 
& $\gamma  n \rightarrow i \pi  N (4 \leq i \leq 8)$\\ 
(35 channels)    &  (35 channels) \\  
                      &           \\ 
\hline 
\end{tabular} 
\label{gat1} 
\end{table} 
\end{centering}

\begin{centering} 
\begin{table} 
\caption{Average numbers of $\pi^+$,$\pi^-$ and $\pi^0$ mesons  
produced in $\gamma A$ and grazing $AA$ collisions} 
\vspace{0.3cm}  
\begin{tabular}{|c|c|c|c|c|c|c|}     
\hline 
  & \multicolumn{3}{c|}{$\gamma A$} & \multicolumn{3}{c|}{AA}  \\   
\cline{2-7} 
  &  &  &  &  &  & \\  
  & $\langle n_{\pi^+}\rangle$  & $\langle n_{\pi^-}\rangle$   
&  $\langle n_{\pi^0}\rangle$ 
  & $\langle n_{\pi^+}\rangle$  & $\langle n_{\pi^-}\rangle$   
&  $\langle n_{\pi^0}\rangle$ \\ 
  &  &  &  &  &  & \\    
\hline 
100A+100A GeV     & 0.25  & 0.31  & 0.35  & 3.15  & 3.21  & 3.26  \\ 
$^{197}{\rm Au}\ on \ ^{197}{\rm Au}$ &  &  &  &  &  &  \\ 
\hline 
2.75A+2.75A TeV     & 0.43 & 0.53 & 0.57 & 10.47 & 10.49 & 10.56 \\ 
$^{208}{\rm Pb}\ on \ ^{208}{\rm Pb}$ &  &  &  &  &  &   \\ 
\hline 
\end{tabular} 
\label{gat2} 
\end{table} 
\end{centering} 
%

\clearpage 
\begin{figure}[ht] 
\begin{center} 
\unitlength=1cm 
\begin{picture}(2.5,2.0) 
\Line(20,100)(60,100) 
\Line(20,90)(60,90) 
\Line(30,105)(40,95) 
\Line(30,85)(40,95) 
\GCirc(60,95){5}{0} 
\Line(60,100)(100,120) 
\Line(60,90)(100,110) 
\Line(75,113)(88,110) 
\Line(85,98)(88,110) 
\Photon(60,90)(70,70){3}{4} 
 
\Line(20,40)(60,40) 
\Line(20,30)(60,30) 
\Line(30,45)(40,35) 
\Line(30,25)(40,35) 
\GCirc(60,35){5}{0} 
\Line(60,40)(100,20) 
\Line(60,30)(100,10) 
\Line(75,18)(88,20) 
\Line(85,33)(88,20) 
\Photon(60,30)(70,60){3}{4} 
 
\GCirc(75,65){5}{0} 
\LongArrow(80,65)(120,80) 
\Line(80,65)(120,50) 
\Line(80,65)(110,40) 
\Line(105,40)(120,40) 
\Line(120,40)(120,55) 
 
\Text(0,3.3)[]{$A_1$} 
\Text(0,1.2)[]{$A_2$} 
 
\Text(2.0,2.7)[]{$\gamma$} 
\Text(2.0,1.8)[]{$\gamma$} 
\Text(4.5,3.0)[]{$\pi$} 
\Text(4.5,1.5)[]{$X$} 
 
 
 
\end{picture} 
\end{center} 
\caption{ 
Meson production in $\gamma\gamma$ fusion.  
\label{NNproc1}  
} 
\end{figure}
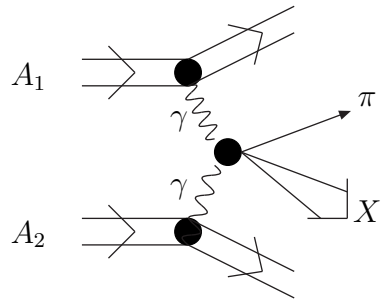 
 
\begin{figure}[h] 
\begin{center} 
\unitlength=1cm 
\begin{picture}(3,4.5) 
\Line(20,100)(60,100) 
\Line(20,90)(60,90) 
\Line(30,105)(40,95) 
\Line(30,85)(40,95) 
\GCirc(60,95){5}{0} 
\Line(60,100)(100,120) 
\Line(60,90)(100,110) 
\Line(75,113)(88,110) 
\Line(85,98)(88,110) 
\Photon(60,90)(70,70){3}{4} 
 
\Line(20,40)(60,40) 
\Line(20,30)(60,30) 
\Line(30,45)(40,35) 
\Line(30,25)(40,35) 
\GCirc(60,35){5}{0} 
\LongArrow(60,40)(100,25) 
\LongArrow(60,35)(100,15) 
\LongArrow(60,30)(100,5) 
\LongArrow(60,30)(70,60) 
 
\GCirc(75,65){5}{0} 
\LongArrow(80,65)(120,80) 
\Line(80,65)(120,50) 
\Line(80,65)(110,40) 
\Line(105,40)(120,40) 
\Line(120,40)(120,55) 
 
\Text(0,3.3)[]{$A_1$} 
\Text(0,1.2)[]{$A_2$} 
 
\Text(2.0,2.7)[]{$\gamma$} 
\Text(2.0,1.8)[]{$N$} 
\Text(4.5,3.0)[]{$\pi$} 
\Text(4.5,1.5)[]{$X$} 
 
 
 
\end{picture} 
\end{center} 
\caption{ 
Meson production in $\gamma A$ collision. 
\label{NNproc2}  
} 
\end{figure}
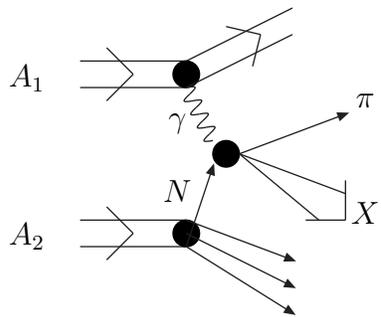 
 
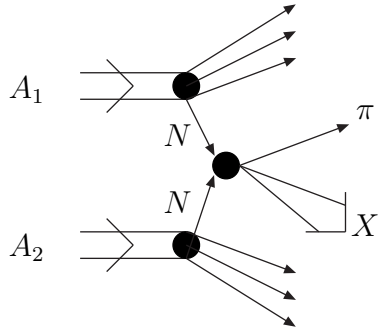
\begin{figure}[h] 
\begin{center} 
\unitlength=1cm 
\begin{picture}(3,4.5)(0,0) 
\Line(20,100)(60,100) 
\Line(20,90)(60,90) 
\Line(30,105)(40,95) 
\Line(30,85)(40,95) 
\GCirc(60,95){5}{0} 
\LongArrow(60,100)(100,125) 
\LongArrow(60,95)(100,115) 
\LongArrow(60,90)(100,105) 
\LongArrow(60,90)(70,70) 
 
\Line(20,40)(60,40) 
\Line(20,30)(60,30) 
\Line(30,45)(40,35) 
\Line(30,25)(40,35) 
\GCirc(60,35){5}{0} 
\LongArrow(60,40)(100,25) 
\LongArrow(60,35)(100,15) 
\LongArrow(60,30)(100,5) 
\LongArrow(60,30)(70,60) 
 
\GCirc(75,65){5}{0} 
\LongArrow(80,65)(120,80) 
\Line(80,65)(120,50) 
\Line(80,65)(110,40) 
\Line(105,40)(120,40) 
\Line(120,40)(120,55) 
 
\Text(0,3.3)[]{$A_1$} 
\Text(0,1.2)[]{$A_2$} 
 
\Text(2.0,2.7)[]{$N$} 
\Text(2.0,1.8)[]{$N$} 
\Text(4.5,3.0)[]{$\pi$} 
\Text(4.5,1.5)[]{$X$} 
 
 
 
\end{picture} 
\end{center} 
\caption{ 
Meson production in grazing  $AA$ collision. 
\label{NNproc3}  
} 
\end{figure} 
 
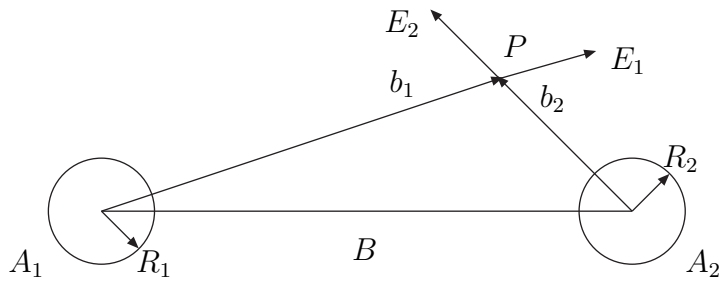
\begin{figure}[h] 
\begin{center} 
\unitlength=1cm 
\begin{picture}(3,3)(2.0,-0.5) 
\CArc(0,0)(20,0,360) 
\CArc(200,0)(20,0,360) 
\LongArrow(0,0)(13,-13) 
\Text(0.7,-0.7)[]{$R_1$} 
\LongArrow(200,0)(213,13) 
\Text(7.7,0.7)[]{$R_2$} 
\Line(0,0)(200,0) 
\Text(3.5,-0.5)[]{$B$} 
\LongArrow(0,0)(150,50) 
\Text(4.0,1.7)[]{$b_1$} 
\LongArrow(200,0)(150,50) 
\Text(6.0,1.5)[]{$b_2$} 
\LongArrow(150,50)(125,75) 
\Text(4.0,2.5)[]{$E_2$} 
\LongArrow(150,50)(185,60) 
\Text(7.0,2.0)[]{$E_1$} 
\Text(-1.0,-0.7)[]{$A_1$} 
\Text(8.0,-0.7)[]{$A_2$} 
\Text(5.5,2.2)[]{$P$} 
\end{picture} 
\end{center} 
\caption{ 
Effective $\gamma\gamma$ luminosity for heavy-ion 
collisions. 
Beam direction is perpendicular to  
the picture plane. $b_1$ and $b_2$ are the distances 
from the nuclear  
centers to the photon interaction point $P$.  
\label{ephoton}  
} 
\end{figure} 

\begin{figure}[hp] 
\begin{centering} 
\epsfxsize=1.1\textwidth 
\epsffile{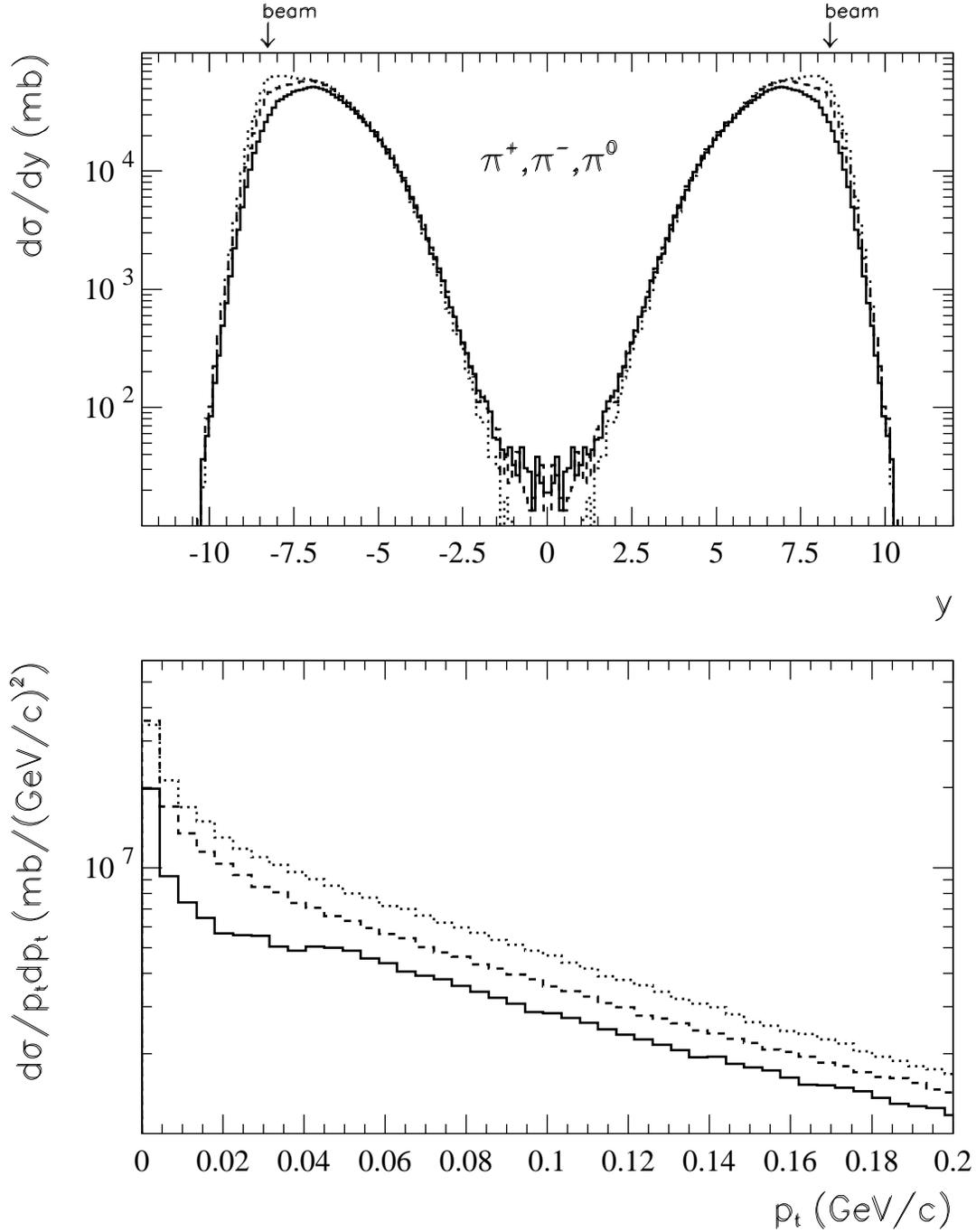} 
\caption{Rapidity and transverse momentum distributions of pions 
produced in $\gamma A$ process in PbPb collisions at LHC. The results  
of RELDIS code are given by solid, dashed and dotted histograms for  
$\pi^+$, $\pi^-$ and $\pi^0$ mesons, respectively. Beam rapidities are 
shown by arrows.} 
\label{gaf1}  
\end{centering} 
\end{figure} 
 
\begin{figure}[hp] 
\begin{center} 
\unitlength=1cm 
\begin{picture}(11,11) 
\put(-2,0){\epsfxsize=14cm \leavevmode \epsfbox{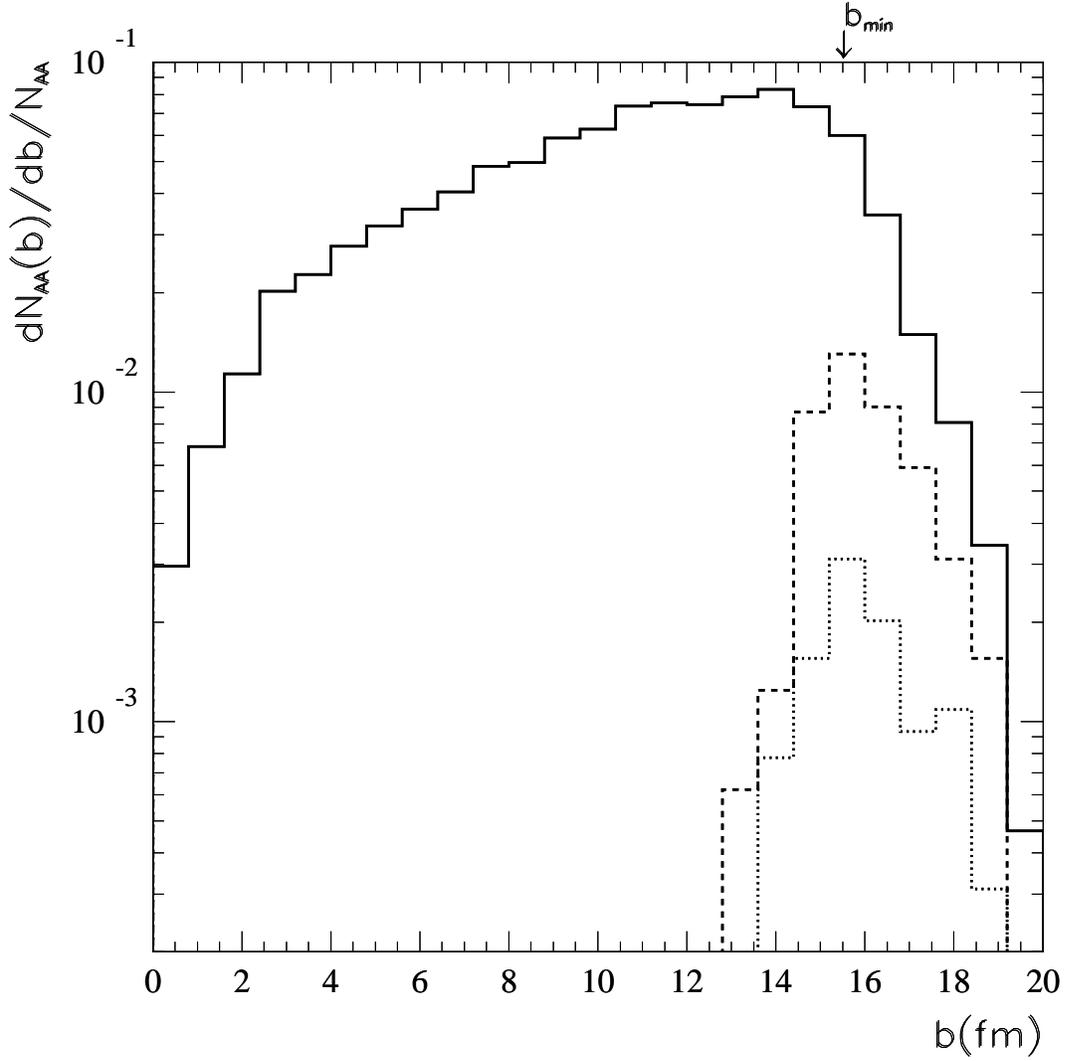}} 
\end{picture} 
\end{center} 
\caption{ 
Impact parameter distributions for nuclear PbPb collisions at LHC
simulated by FRITIOF code. 
The solid histogram corresponds to all events.
The doted and dashed histograms present the impact parameter distributions
for events with $N_{\pi^0}=1$ and $N_{\pi^0}=2$, respectively. 
\label{impact0}  
} 
\end{figure} 
 
\begin{figure}[hp]
\begin{center}
\unitlength=1cm
\begin{picture}(11,11)
\put(-2,0){\epsfxsize=14cm \leavevmode \epsfbox{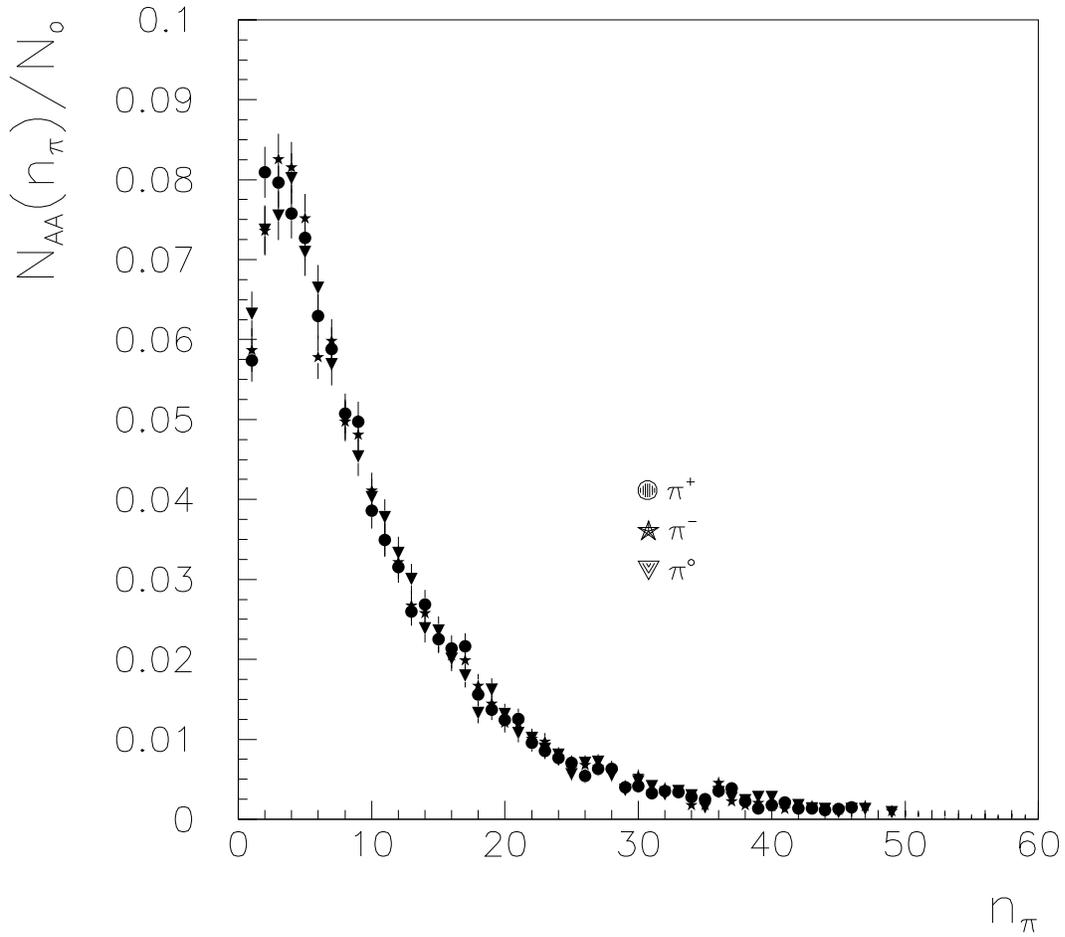}}
\end{picture}
\end{center}
\caption{
Pion multiplicity distributions for grazing nuclear PbPb collisions at LHC simulated 
by FRITIOF code. The events with impact parameter $b\ge 15.5$~fm were selected.
\label{N_n_tot} 
}
\end{figure}

\begin{figure}[hp] 
\begin{centering} 
\unitlength=1cm 
\begin{picture}(11,11) 
\put(1,0){\epsfxsize=14cm \leavevmode \epsfbox{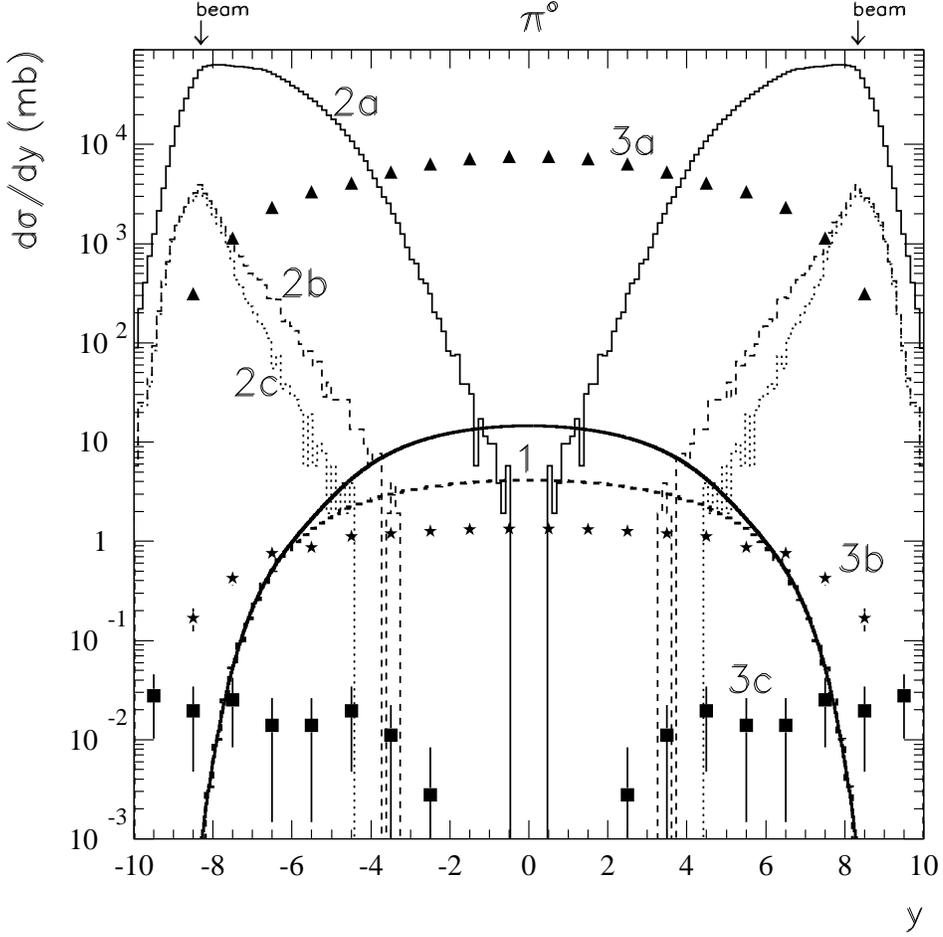}} 
\end{picture} 
\end{centering} 
\caption{ 
Rapidity distribution of $\pi^0$ produced in
peripheral PbPb collisions at LHC energies. 
The thick solid and dashed line histograms labeled ``1'' are the inclusive 
(DRP model) and 
exclusive cross sections for $\pi^0$ production in $\gamma\gamma$ fusion. 
The histograms ``2a'',``2b'' and ``2c'' give the results of RELDIS 
code for $\gamma A$ process.
The FRITIOF results for peripheral PbPb collisions 
with $b\geq 15.5$~fm are given
by the points ``3a'',``3b'' and ``3c''.
The distributions with label ``$a$'' were obtained  without $p_t$ cuts. 
The label ``b'' corresponds to the selection of events 
with the total transverse momentum of meson system 
$|\vec {p_t}^{(sum)}|\leq 75$~MeV/c.
The label ``c'' corresponds to the selection of events  
with $|\vec {p_t}|\leq 75$ MeV/c for each of the pions 
including $\pi^0$, $\pi^+$, and $\pi^-$.
\label{incl0}  
} 
\end{figure} 
 
\end{document}